# Deep-Learning-Based, Multi-Timescale Load Forecasting in Buildings: Opportunities and Challenges from Research to Deployment

Sakshi Mishra[†], Stephen M. Frank, Anya Petersen, Robert Buechler, and Michelle Slovensky

Intelligent Campus Program, National Renewable Energy Laboratory, Golden, Colorado, United States

[†]Corresponding author (sakshi.mishra@nrel.gov)

*Abstract*— Electricity load forecasting for buildings and campuses is becoming increasingly important as the penetration of distributed energy resources (DERs) grows. Efficient operation and dispatch of DERs require reasonably accurate predictions of future energy consumption in order to conduct near-real-time optimized dispatch of on-site generation and storage assets. Electric utilities have traditionally performed load forecasting for load pockets spanning large geographic areas, and therefore forecasting has not been a common practice by buildings and campus operators. Given the growing trends of research and prototyping in the grid-interactive efficient buildings domain, characteristics beyond simple algorithm forecast accuracy are important in determining the algorithm's true utility for smart buildings. Other characteristics include the overall design of the deployed architecture and the operational efficiency of the forecasting system. In this work, we present a deep-learning-based load forecasting system that predicts the building load at 1-hour intervals for 18 hours in the future. We also discuss challenges associated with the real-time deployment of such systems as well as the research opportunities presented by a fully functional forecasting system that has been developed within the National Renewable Energy Laboratory's Intelligent Campus program.

## I. INTRODUCTION AND BACKGROUND

The commercial buildings sector in the United States consumed 1,671.61 trillion Btu of energy in January 2020 alone (immediately prior to the onset of the COVID-19 pandemic) and a total of 18,177.95 trillion Btu in 2019 (Total Energy - Monthly Energy Review, 2020). Overall, buildings account for nearly 40% of the total energy consumption in the United States (DOE), and building energy consumption is projected to expand by an annual 1.5% globally (Jordan, 2014). Even with many commercial facilities operating at reduced occupancy during the COVID-19 pandemic, preliminary industry reports indicate that commercial building energy is again increasing as operators increase ventilation rates (Energy-Spectral, 2020). Because buildings are among the largest consumers of energy globally, research to increase the design and operational energy efficiencies of the commercial sector (i.e., commercial buildings) will play a significant role in meeting energy and greenhouse gas emissions reduction targets.

The penetration of behind-the-meter distributed energy resources (DERs) is also increasing, owing to the decreasing cost of renewable energy technologies (Donohoo-Vallett, et al., 2016). Smart buildings operations can simultaneously reduce energy consumption (and, by extension, greenhouse gas emissions) and optimize behind-the-meter DER dispatch to save money for building owners/operators via additional value streams such as demand management (peak shaving), energy arbitrage, user-initiated demand-response, and optimal electric vehicle charging. Properly deployed, these capabilities also benefit the utility grid by enhancing grid reliability and resilience, deferring or reducing capital expenditures required to upgrade the distribution grid, and helping balance the supply of renewable energy as its penetration increases. The U.S. Department of Energy's Grid-interactive Efficient Buildings initiative (Neukomm, Nubbe, & Fares, 2019) describes the multi-faceted benefits of these smart building capabilities. Smart building controls are also an integral component of autonomous energy grids (Kroposki, Dall'Anese, Bernstein, Zhang, & Hodge, 2017), where they provide opportunities for granular device-level controls.

Proportional–integral–derivative (PID) controllers are still widely used in building controls for heating, ventilating, and air-conditioning (HVAC) systems, largely because of their simplicity and fast solutions times as they employ numerical methods to determine the controlling parameters (Afram & Janabi-Sharifi, 2014). However, the time-varying system dynamics characteristics of HVAC control systems result in the inconsistent performance of traditional controllers, often leading to less-economical performance (Khanmirza, Esmaeilzadeh, & Markazi, 2016). Moreover, the objectives of building control systems have shifted from being solely occupant-comfort and economic-operations to being grid-interactive prosumers. As prosumers, building energy management systems have goals such as (i) maximize the use of on-site distributed energy resources, (ii) manage consumption profile based on time-of-use energy rates, and (iii) use on-site energy storage to reduce the peak demand for cost savings.

Such a complex set of objectives requires the use of advanced control techniques like model-predictive control (Khanmirza, Esmaeilzadeh, & Markazi, 2016). Implementation of intelligent dispatch and model-predictive control algorithms that help achieve the above-stated goals to reduce overall energy consumption and operational costs of the building nearly always requires a forecast of the energy consumption of the building. These algorithms use these building load predictions to schedule the dispatch of flexible loads, on-site clean energy resources, and/or energy storage systems to optimize a desired metric, such as utility cost. Forecast accuracy is a key driver of model-predictive control algorithm effectiveness (Afram & Janabi-Sharifi, 2014) (Khanmirza, Esmaeilzadeh, & Markazi, 2016) (Nguyen, Yoo, & Kim, 2017). Prediction of building energy consumption patterns is also important for detecting faults or operational anomalies in energy systems. Therefore, building load forecasting systems are an essential component of smart buildings.

Building load forecasting is not a new research question; many research works in the literature have addressed this problem space (Iino, et al., 2009) (Zhang & Zhou, 2008) (Iwafune, Yagita, Ikegami, & Ogimoto, 2014). Commercial load forecasting tools for building energy predictions are also available (Itron) (Enverus). The journey of successful research outcome to industry deployment, however, inevitably passes through the research-prototyping phase, and advanced building load forecasting systems are no exceptions. This is especially relevant for data-driven modeling projects where the type, quality, and preprocessing of data largely dictate the performance of the forecasting system. The work presented in this article sheds light on the end-to-end pipeline of building load forecasting systems—in a campus-wide prototyping setup—using advanced data-driven methods (recurrent neural networks). The following paragraphs delineate the types of modeling algorithms used in the literature for the building energy forecasting problem and specifically discuss works that have utilized data-driven modeling methods.

As building systems become more advanced through next-generation sensors, controls, connectivity, and communications, they produce a large volume of empirical data available to building operators for decision-making. These data, along with meteorological parameters, can be harnessed to predict building and campus energy consumption. There are two main approaches available for this prediction task: (i) physics-based (or "white box") modeling, and (ii) data-driven (or "black box") modeling. (Physics-based models are termed "white box" because the inner workings of the model are typically open to the modeler, whereas data-driven models are termed "black box" because their prediction logic is often opaque.) A third category, "grey box" models (Sohlberg & Jacobsen,

2008), represent a combination of the physics-based and data-driven approaches. They combine a partial theoretical structure with data, offering relatively simpler model architectures. Grey box models are generally trained or fit like data-driven models.

Physics-based building models such as EnergyPlus® (Crawley, et al., 2001), (EERE, 2020) model the physical relationships between the building characteristics (construction details, operation schedules, shading information) and environmental parameters (sky conditions) to calculate building energy consumption (Kwok & Lee, 2011). Data-driven models, on the other hand, make predictions by learning the pattern empirically from historical data. Data-driven models have two major subcategories: (i) statistical models, and (ii) machine learning models. Deep learning models are a subset of machine learning models that are capable of learning complex nonlinear relationships between the inputs and the predicted variable(s). Recurrent neural networks (RNNs) are an advanced variant of deep neural networks that are capable of incorporating temporal dependencies between the input and output variables. Long short-term memory networks (LSTM) are a variant of RNNs that are effective at capturing longer-term temporal dependencies in the data sets.

There are numerous advanced applications of predictive analytics in the renewable energy field (Mishra, Glaws, & Palanisamy, 2020)—spanning the energy system from generation (solar and wind forecasting) to consumption (smart buildings energy forecasting to fault predictions). Because the focus of this article is smart buildings' predictive analytics application, herein we discuss literary works focused on data-driven building load forecasting. Recent research has established the utility of neural networks for forecasting the energy consumption of both individual buildings and groups of buildings. For example, a machine-learning-based forecasting model is presented in (Hwang, Suh, & Otto, 2020); whereas (Jahan, Snasel, & Misak, 2020) covers a review of the intelligent system for power load forecasting. A Polish power system study with a deep learning approach is presented in (Ciechulski & Osowski) and (Jin, Guo, Wang, & Song, 2020) takes a deeper look at a hybrid system based on LSTM for short-term power load forecasting; similarly, (Chitalia, Pipattanasomporn, Garg, & Rahman, 2020) presents an RNN-based robust short-term load forecasting framework. An online adaptive RNN-based load forecasting algorithm with smart meter data is presented in (Fekri, Patel, Grolinger, & Sharma, 2021). Short-term load forecasting for urban loads using an artificial neural network is tackled in (Gautam, Mayal, Ram, & Priya, 2019). An LSTM based method is improved upon for short-term load forecasting in (Cui, Gao, & Li, 2020).

On the research end of the building load forecasting spectrum, the existing literature has extensively focused on assessing the effectiveness of machine learning (specifically deep learning) models for the building load prediction problem. The commercial end of building load forecasting spectrum, on the other hand, focuses on using proven and tested traditional methods in prepackaged software products. Both these extreme ends of the building load forecasting problem space do not address the "research prototyping" step for incrementally adopting cutting-edge prediction methodologies. That is to say, none of these works have covered a holistic picture of the deep-learning-based prediction system deployment process—in a research set-up—which starts with data collection, curation process, and continues to critically assess the challenges and opportunities with its end-to-end deployment and continued use. The work presented in this paper contributes a perspective of how to move data-driven algorithms from research to practical deployment by presenting a case study of designing, building, and deploying a research prototype of a deep-learning-based building load forecasting system.

The current section lays out the introduction and background behind the presented work, and the rest of the paper is organized as follows: Section II provides a brief introduction of the real-time testbed (Intelligent Campus) utilized for deploying the deep learning models. Section III discusses the algorithm employed in modeling, methodology, and benchmarking. Section IV presents the challenges faced, lessons learned, and opportunities that a fully functional and deploy load forecasting system presents for modern buildings. Section V concludes the paper with an outlook discussion.

## II. INTELLIGENT CAMPUS

The Intelligent Campus platform at the National Renewable Energy Laboratory (NREL) was established to collect historical and real-time building performance data to support analytics that enhance operational awareness and decision-making with respect to energy use. The platform was developed using open standards and protocols (Fig. *1*) with the intent to provide an architecture that is readily transferable to other campuses (Cutler, Frank, Slovensky, Sheppy, & Petersen, 2016). Since its inception, the Intelligent Campus platform has evolved into a program providing an ecosystem for the use of the NREL campus as a living laboratory.

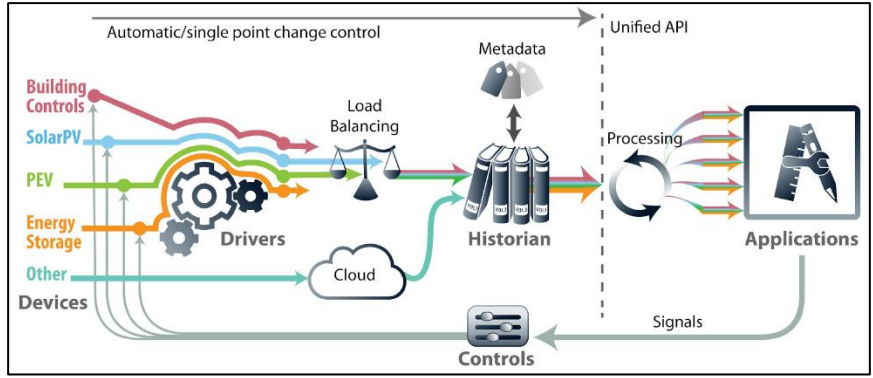

Fig. 1. End-to-end data collection architecture

The Intelligent Campus program offers nascent energy analytics technologies and algorithms a pathway to mature toward commercialization. The Intelligent Campus program includes an interdisciplinary portfolio of projects aimed at prototyping the future of smart, sustainable, resilient, and self-healing buildings. By providing an environment for technology demonstration in real facilities, the Intelligent Campus program is bridging the gap between theoretical research into building efficiency technologies (which is often published but rarely deployed) and practical application. Intelligent Campus pilot projects include detailed monitoring and analysis in order to identify and improve technology shortcomings.

*A. Data: Measurement, Collection, and Curation*

Advanced analytics and model-predictive control algorithms for smart buildings require reliable, internally consistent building performance data. NREL collects a variety of performance data at its South Table Mountain campus. The campus has a single common utility electric meter for all facilities, including the campus's central heating and chilled water plant. Eighteen campus buildings have whole-building electrical meters; several other small buildings do not have dedicated electrical meters. Several newer buildings have submetering by end use, per the requirements of ASHRAE 90.1 (ASHRAE, 2019). Facilities that consume heating and chilled water from the central plant have thermal meters for heating and chilled water consumption.

NREL captures the electric and thermal meter interval data using a central energy management information system (EMIS). The EMIS also collects interval data from the campus building automation system and syncs data from several cloud data sources via web application programming interfaces, including utility data from the campus main electricity meter and NREL's on-site research weather station (Stoffel & Andreas, 1981). To maximize accessibility for operations and research, Intelligent Campus organizes collected interval data per the Project Haystack standard (Project Haystack, 2020) (Cutler, Frank, Slovensky, Sheppy, & Petersen, 2016). These interval data are available to serve as inputs for predictive load models.

*B. Predictive Analytics*

Intelligent Campus's predictive analytics project applies recent machine learning and deep learning advances to develop and continuously improve load forecasting capability for NREL's South Table Mountain campus. Intelligent Campus executes research, prototyping, and enhancement in a cycle so that practical deployment lessons can be folded back into early-stage research, increasing the ultimate effectiveness and eventual impact of the work. Intelligent Campus's building load forecasting work builds from prior NREL research that predicts expected building performance based on exogenous inputs (Henze, Pless, Petersen, Long, & Scambos, 2015).

As an academic discipline, load forecasting approaches can be studied using static postprocessed data sets. Transferal of these results to a real-world setting requires at a minimum some underlying sensor infrastructure, a mapping of sensor data streams to model inputs, and quality control on the measured data being fed to the model. Therefore, the focus of the current effort is twofold: (i) to automate, replicate, and scale the predictive algorithms; and (ii) to improve accuracy by leveraging state-of-the-art neural networks.

### III. LOAD FORECASTING USING DEEP LEARNING

*A. Deep Learning: Long Short-Term Memory Network*

Artificial neural networks are universal function approximators. They are capable of representing complex nonlinear relationships in high-dimensional data sets such as the one being employed for building load forecasting in this work. Deep learning algorithms, which employ multiple hidden layers, have powerful generalizing capabilities. That is, they are able to make reasonably accurate predictions for previously unseen scenarios because of their ability to learn the intricate structures in large data sets. Feedforward neural networks, sometimes referred to as "vanilla" deep neural networks, have the fundamental drawback of independence among the time-series samples/data points, which makes them an ineffective choice of algorithm for time-series-based prediction problems. This is because the entire state of the feedforward neural network is cleared after processing each time-series sample, which means that the network starts mapping the inputs and output for the next time step from scratch, thereby failing to account for the impact of the previous time step's inputs variables.

Recurrent neural networks (RNN) are advanced types of deep neural networks that overcome feedforward neural networks' limitations. They are different from feedforward neural networks because of the presence of additional directed edges that introduce temporal memory components, enabling them to capture complex nonlinear relationships between the temporally related inputs and outputs across multiple time steps. This makes them effective at modeling time-series forecasting problems. A long short-term memory (LSTM) network (Hochreiter & Schmihuber, 1997), shown in Fig 2, is an RNN variant that is able to learn *long-term* dependencies between the input features and the predicted variable (Mishra & Palanisamy, 2019).

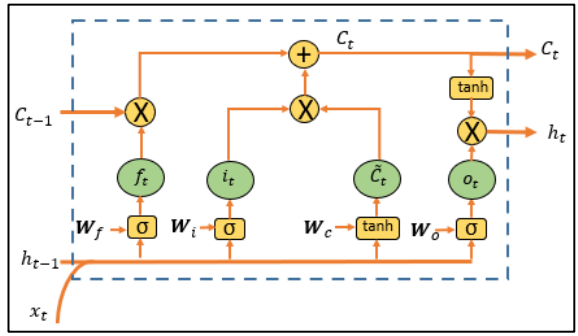

Fig 2. LSTM memory cell diagram[1] *(Mishra & Palanisamy, 2019)*

Fig 2 displays a typical memory cell for an LSTM network. In the figure, $C$ denotes the cell states at different points in time $t$; $f_t$, $i_t$, and $o_t$ denote the forward, input, and output gates, respectively; $h^{(t)}$ is the hidden state at time $t$; and $W_f$, $W_i$, $W_c$, $W_o$ are the weights. Further explanation can be found in (Mishra & Palanisamy, 2019).

*B. Performance Measure*

The evaluation metric used to measure the performance of the forecasting algorithm is the mean squared error (MSE). The difference between the ground truth (i.e., the actual future time-step value of the measurement being predicted) and forecasting values (i.e., output of the deep learning model) is calculated using MSE:

$$\text{MSE} = \frac{1}{n} \sum_{j=1}^{n} \left( AV_j^{(t)} - \widehat{OV_j^{(t)}} \right)^2$$

where $AV_j^{(t)}$ is a vector of actual (ground truth) values, and $\widehat{OV_j^{(t)}}$ is a vector of forecasted (output) values.

*C. Case Study: Building Load Forecasting on a Research Campus*

In this case study, the energy consumption at the main electricity meter for the Café building on the NREL South Table Mountain campus is predicted at an hourly resolution. There are six input features used for training the model: relative humidity, barometric pressure, dry bulb temperature, global horizontal irradiance, total cloud cover, and wind speed. These data are from NREL's Solar Radiation Research Laboratory data set (Stoffel & Andreas, 1981); however, they also represent measurements commonly available from high-quality weather stations worldwide. The model is run for 200 epochs with the data spanning a year's interval: 10- and 2-months train/test split, respectively. We use a single-layered network with 35 neurons. The following MSE plot shows the train and test losses over several thousand iterations. Despite the difference between final train and test losses, no overfitting was observed in the experiment.

---

[1] Reused with permission

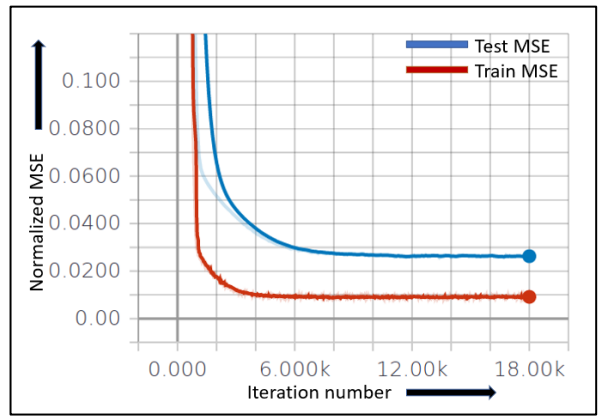

Fig 3. Mean squared error (MSE) error plot for train and test data

The model can produce load forecasts at multiple future time steps, as shown in Fig 4 for the Café building main power. The models use the most recent 12 hours of weather data to capture the transient effect of weather on energy consumption and use time of day and day of week to capture time-based energy consumption patterns, such as those related to building occupancy. Because neural networks can learn non-linear relationships, the trained models are able to infer complex behaviors, such as the pre-occupancy energy spike in the Café due to early morning food preparation. Each model outputs a prediction of building load from the current time step to 18 hours in the future in 1-hour increments. This input-output structure applies to all buildings that were studied on the NREL campus.

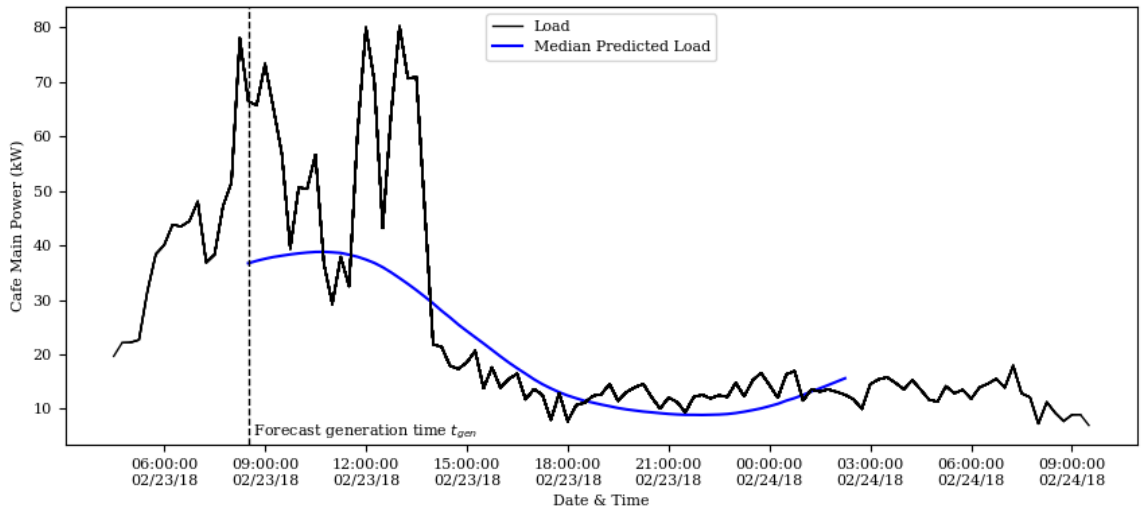

Fig 4. Multi-time horizon forecast

Model performance was not seen to change significantly as the length of the forecast window increased/decreased, but the training time was significantly impacted by the length of the training window (number of hours of historical data used as input to the model).

## IV. OPERATIONAL LOAD FORECASTING SYSTEM DEPLOYMENT

A primary research objective of the case study was to establish deployment and continuous improvement pathways for the developed algorithms. That is, developed algorithms should be able to be rapidly deployed for testing within the Intelligent Campus platform and iteratively improved with minimal retooling. The Intelligent Campus team intends to expand the algorithm to forecast other quantities (such as building electricity meters, thermal meters, and photovoltaic system generation) and to support additional exogenous variables. Therefore, from a software development point of view, the system's input-output architecture must be flexible, replicable, and scalable.

Due to the nature of the deep learning algorithms used, ongoing access to data is an integral part of the continued operation of the system. Periodic retraining of the models as more data gets collected has the potential to increase model accuracy. The following two subsections describe the challenges and opportunities that open up for smart buildings' research and prototyping once those challenges are addressed. Understanding the deployment aspect of deep-learning-based forecasting system, therefore, completes the research-prototyping loop.

### A. Challenges

#### 1) Data Availability and Quality Control

"Garbage in, garbage out"—a common phrase used in the machine learning community—captures the importance of this step of data collection and curation. Assembling a machine learning training data set begins with proper commissioning and maintenance of meters, sensors, and other data inputs. Data should have proper units and scaling. (Although in some circumstances a machine learning model could be successfully trained from improperly scaled or otherwise corrupted input data, if those data stream errors were corrected in the future the model would need to be retrained.)

Automated accumulation of new training data (without extensive human review) presents data quality challenges. NREL's Intelligent Campus team has implemented quality check mechanisms for the measurements coming from the sensors in real-time before they are fed to the model. One example is removing the outliers identified by the values lying outside of three standard deviations. Another is operating basic fault detection rules that check for stuck sensors (constant value for long periods of time), logically out-of-range data (such as significant solar radiation recorded at night), and similar anomalies. These checks not only improve machine learning model quality but also provide an opportunity to detect and correct errors in the measurement systems and equipment.

#### 2) Integration with EMIS and Scalability

Practical implementation of a forecasting system requires reliable interfaces for transferring data to the model and for transferring model outputs to an analytics system that consumes them. The forecasting system developed for this case study communicates with the NREL EMIS via an application programming interface defined by the Project Haystack standard (Project Haystack, 2020). The EMIS provides forecast inputs, and the forecast system, in turn, writes forecasts back to the EMIS. Adherence to a metadata standard and an open application programming interface allows the forecasting system to be rapidly retrained and redeployed for new sites. Any database that adheres to the Project Haystack standard and correctly tags the required input points can be connected to a new instance of the deep learning algorithm, and a new model can be trained with minimal effort. When the forecasting system runs in integration with EMIS, it is also important to be able to redeploy the model (with either architectural or training updates) without service interruption.

Because forecasts cover a span of time but are frequently updated, managing caching is a key challenge. In addition, the forecast read/write system must be flexible enough to handle arbitrary forecast inputs and outputs. Additionally, given the computationally intensive nature of deep learning models, scalability in terms of available compute power for retraining the models is another interesting challenge.

*3) Online vs. Offline Training*

Training is an important consideration for deep-learning-based forecasting systems because deep learning models' accuracies have been shown to keep rising with more training. Offline training is a relatively simpler way of implementation in which the machine learning models are retrained by a manual process where an engineer copies the old model to a local machine, trains it further with the new data, and transfers it back to the servers where the systems are deployed. Offline training, therefore, requires more human-hours over the course of the system's operation because human intervention is needed to further train the models at regular intervals.

The online training approach, on the other hand, is an automated process. At its simplest, an online training system trains models on a schedule (for example, every 1 to 2 months), without any manual intervention (Cloud-Architecture-Center, Google, 2021). Though setting up the automated training pipeline initially requires additional planning and coding, this online training eliminates the maintenance dependency on human intervention to move the models offline, retrain them, and upload them back into the forecasting system pipeline regularly. Thus, online training can increase operational efficiency and reduce opportunities for human error to impact the forecasting system. Moreover, because a robust metering platform is already established as the first step in the process in our prototype of Intelligent Campus, high-quality data are constantly accumulating and can be used for increasing the accuracy rates of our models. For our final implementation, we have chosen the online training option.

*4) Continuous Improvement*

As described above, deep learning models are capable of increasing accuracy as more data is made available for training. Also, in the initial stages when the accuracy of the model is not satisfactory given the shortage of data points to train it, large changes in predictive behavior can be off-putting to end users. Therefore, along with iterative training, we conduct model enhancement with hyperparameter tuning as the operational circumstances change over time. This is an infrequent yet desirable step.

*5) Generalizability*

Another typical challenge in deploying deep learning models for multiple prediction points (i.e., building wise energy consumption forecasts) is generalizability. Because each prediction point has its own model, the deployment architecture must address the need for automated model generation. To address the challenge of generalizability, the forecasting system's data preprocessing and postprocessing modules are designed with the capability to take in a generic variable named "point-id" that is used as a reference to fetch the respective machine learning model and input data streams.

*B. Opportunities*

An automated end-to-end pipeline for load forecasting deployed on-site, with an online training mechanism in place, enables many opportunities to harness the predicted data for furthering the smart buildings' research. In the following subsections, we describe three potential applications.

*1) Demand Management*

Motivations for electricity demand management (peak shaving) including utility demand charge reductions and reduced carbon footprint. (When the peak of multiple buildings and campuses coincides with the utility peak, lower-efficiency "peaking" power plants fired by gas or diesel are used to support the grid, which translates to proportionally higher greenhouse gas emissions.) Reducing peak demand requires the ability to predict the timing and magnitude of peak demand, then preemptively shifting load or dispatching energy storage assets to avoid the peak.

*2) Energy Arbitrage*

Energy arbitrage is another cost-saving opportunity available for grid-interactive buildings in cases where on-site energy storage is available. Energy arbitrage is performed by participating in the energy markets by charging energy storage (for example, batteries) when electricity prices are low and discharging storage to sell power back to the grid when prices spike. Even in utility markets where bidirectional electricity exchange is not possible, a form of energy arbitrage is possible by buying energy when prices are low, storing the energy locally, and dispatching the energy locally to meet load when energy prices are high.

Energy arbitrage can be accomplished by employing near-real-time or real-time optimization algorithms. However, the real-time optimized storage dispatch requires reasonable foresight into future on-site energy consumption, thereby making building load forecasts an important component of the system.

*3) Load Flexibility*

A third opportunity to save energy costs for building operations is shifting the load based on the time-of-use energy rates. Building load can be classified as critical, which consists of loads that cannot be deferred, and flexible, which consists of loads whose consumption can be scheduled within a given time window. Examples of critical loads include lights, fans, microwaves, computing systems, etc.; these loads must be available for use when needed. Examples of flexible load include dishwashers, laundry, and even heating and cooling (buildings can be precooled/preheated). With adequate foresight, flexible loads can be shifted to minimize the energy costs. This also requires predicting the building's future energy consumption.

## V.  SUMMARY AND OUTLOOK

The paradigm of grid-interactive efficient buildings, i.e., dynamically operating buildings that constantly exchange information and energy with the utility grid, is increasingly gaining traction in both the research community and in industry. Such smart buildings will operate in harmony with the grid to make electricity more affordable and integrate a larger share of DERs while meeting the comfort and productivity needs of the buildings' occupants. NREL's Intelligent Campus is a living laboratory that is dedicated to accelerating the research and deployment of various pieces of the smart buildings puzzle, serving as a vital testbed for the innovative solutions that push the frontier of smart building operations.

In this article, we describe and critically discuss a deep learning building load forecasting system that is an important module of smart buildings' overall controls. We demonstrate that metering, collection, and curation of data is a crucial part of the prototyping process of such an advanced forecasting system. Using LSTMs, the deep learning model can predict the building and campus load for 18 hours' time-horizon. The architecture is flexible with the time-resolution; subhourly predictions can be enabled with minimal modifications to the code.

Using Intelligent Campus as the platform for testing the research findings in real-time settings, we shed light on the deployment aspect of such advanced forecasting systems. We conclude that a well-architected design of the forecasting system is key to its effectiveness. To take a forecasting algorithm from the research phase to deployment for practical use, building engineers will need to focus on the end-to-end pipeline, which requires considerations for multiple submodules such as access to data, data handling, seamless integration with the existing building data platform, model training and update mechanism, periodic performance checks, and model enhancement with hyperparameter tuning as operational circumstances change.

The end-to-end forecasting system presented in this work forms a robust platform for furthering research in the building energy predictive analytics problem space. We plan to expand this work in the following directions:

*A.   Probabilistic Forecasting*

Rather than generating a point forecast for every time step, a range can be outputted with the estimated minimum and maximum values serving as the upper and lower bounds for a prediction band within which the actual consumption in the future time step will fall. Having a range instead of point forecasting is especially helpful in the fault detection application where consistent outliers (outside of the band points) indicate a potential issue which can be further investigated by the building's operations team. Moreover, probabilistic forecasting using deep neural networks is an active inquiry-front on the research end of the forecasting technology's spectrum.

*B.   Predictive Maintenance*

Current practice in the buildings industry is to perform "schedule-based" maintenance for fault prevention, which is not effective in terms of flagging the equipment vulnerable to a fault beforehand. The next generation of the fault-prevention mechanism is "condition-based" or "preventive" maintenance. This involves using statistical analysis to assess the health and estimate the probability of failure, to inform the operators. The presented forecasting system can be further developed for predictive-maintenance application where the occurrence of a specific fault on the system, which may occur due to unusual wear-and-tear on equipment can be predicted well in advance, thereby preventing equipment failure triggering faults.

ACKNOWLEDGMENT

This work was authored by NREL, operated by Alliance for Sustainable Energy, LLC, for the U.S. Department of Energy (DOE) under Contract No. DE-AC36-08GO28308. Funding provided by the U.S. Department of Energy Office of Energy Efficiency and Renewable Energy Building Technologies Office. The views expressed in the article do not necessarily represent the views of the DOE or the U.S. Government. The U.S. Government retains and the publisher, by accepting the article for publication, acknowledges that the U.S. Government retains a nonexclusive, paid-up, irrevocable, worldwide license to publish or reproduce the published form of this work, or allow others to do so, for U.S. Government purposes.

The authors wish to acknowledge the National Renewable Energy Laboratory's continued support in enhancing the research and prototyping capabilities of the Intelligent Campus program. The authors would like to thank Rob Buechler (Stanford University) for reviewing the code implementation.